\documentclass{aa}
\usepackage{amsmath} 
\usepackage{graphicx}
\usepackage{natbib}
\bibpunct{(}{)}{;}{a}{}{,}
\usepackage{txfonts}
\usepackage[T1]{fontenc}
\usepackage{ulem}
\usepackage{xcolor}
\newcommand{\cmd}{\,cm$^{-2}$}

\newcommand{\myr}{\,$M_{\odot}\,{\rm yr}^{-1}$}
\newcommand{\ro}{\,$R_{\odot}$}
\newcommand{\mo}{\,$M_{\odot}$}
\newcommand{\lo}{\,$L_{\odot}$}
\begin{document}

%
%
\title{Wind-mass transfer in S-type symbiotic binaries}
\subtitle{IV. Indication of high wind-mass-transfer efficiency 
          from active phases}
\titlerunning{Wind-focusing in S-type symbiotic binaries}
\authorrunning{A.~Skopal, N.~Shagatova}
\author{A.~Skopal, \and N.~Shagatova}
\institute{Astronomical Institute, Slovak Academy of Sciences,
        059~60 Tatransk\'{a} Lomnica, Slovakia
        \email{skopal@ta3.sk, shagatova@ta3.sk}
}
\date{Received / Accepted }

\abstract
{
Observational indications of wind-mass transfer from an evolved 
giant to its distant white dwarf (WD) companion in symbiotic 
binaries are rare. 
Here, we present a way to examine the neutral wind 
from the giant in symbiotic binaries, which is temporarily 
observable throughout the orbital plane during outbursts. 
}
{
We find that the mass-loss rate from giants in the orbital 
plane of S-type symbiotic binaries is high, indicating a high
wind-mass-transfer efficiency in these systems. 
}
{
We modeled hydrogen column densities in 
the orbital plane between the observer and the WD for all 
suitable eclipsing S-type symbiotic binaries during outbursts 
in any orbital phase.  
}
{
The mass-loss rate from the giant in the 
orbital plane is on the order of 10$^{-6}$\myr, which is 
a factor of $\sim$10 higher than rates derived from nebular 
emission produced by the ionized wind from normal giants in 
symbiotic stars. This finding suggests a substantial focusing 
of the giant's wind toward the orbital plane and, thus, its 
effective transfer onto the WD companion. 
}
{
Our finding suggests that wind focusing on the orbital 
plane may be a common property of winds from giants 
in S-type symbiotic stars. 
Such wind-focusing resolves a long-standing problem of the large 
energetic output from their burning WDs and deficient fueling 
by the giant via a standard Bondi-Hoyle accretion. It also allows 
the WD to grow faster in mass, which lends support to the possibility 
that S-type symbiotic binaries are progenitors of Type Ia 
supernovae. 
}
\maketitle
\section{Introduction}
\label{s:intro}
Symbiotic stars (SySts) are the widest interacting binaries; their 
orbital periods are typically between 1 and 3 years 
\citep[for S-type systems that contain a normal giant as the donor; 
see][]{2000A&AS..146..407B,2013AcA....63..405G} but can be 
significantly longer for systems that contain a Mira variable 
\citep[D-type systems;][]{
1975MNRAS.171..171W,2013ApJ...770...28H}. 
The accretor is a compact star, most often a white dwarf (WD). 
Accordingly, the large separation of their components suggests 
that SySts are detached binaries 
\citep[e.g.,][]{1999A&AS..137..473M}, that is to say, their activity 
is triggered via wind-mass transfer\footnote{Roche-lobe overflow 
in SySts has not yet been proven; see Sect.~\ref{ss:dis2}.}. 

%
The accretion process is responsible for the WD's temperatures 
of $1-2\times10^5$\,K and luminosities of $10^{1}-10^{4}$\lo\ 
\citep[e.g.,][]{1991A&A...248..458M,2005A&A...440..995S}. 
Depending on the accretion rate and the WD mass, the accreting 
material can either convert only its gravitational potential 
energy into radiation through the accretion disk and 
the boundary layer -- the so-called accretion-powered 
systems\footnote{Sometimes referred to as accreting-only SySts 
\cite[][]{2019arXiv190901389M}.} 
with luminosities of a few times 10$^{1}$\lo\ 
\citep[e.g.,][]{2005ASPC..330..463S,
                2016MNRAS.461L...1M,
                2021MNRAS.505.6121M} 
-- or it can also nuclearly burn on the WD surface in stable 
conditions, when the hydrogen-rich material is burning as it is 
accreting -- the so-called nuclear-powered systems,\footnote{Also known 
as as shell-burning systems \citep[e.g.,][]{2016acps.confE..21S}.}
which generate luminosities of a few times 10$^{3}$\lo\ 
\citep[e.g.,][]{
  1978ApJ...222..604P,
  2007ApJ...660.1444S}. 

We distinguish 
between quiescent and active phases of SySts according to the optical 
light variations. 
During quiescent phases, the accreting WD ionizes the neighboring 
part of the giant's wind, giving rise to the nebular emission 
\citep[e.g.,][]{1966SvA....10..331B,1984ApJ...284..202S}. 
Due to the high wind densities in the binary, the nebula is 
partially optically thick, which causes periodic wave-like 
variations in light curves as a function of the orbital phase 
\citep[][]{2001A&A...366..157S}. 
During active phases, unstable nuclear burning on the WD 
surface gives rise to transient outbursts indicated by 
brightening of a few magnitudes in the optical. 
The nuclear-powered systems can give rise to the so-called 
Z~And-type outbursts that result from an increase in the 
accretion rate above that sustaining the stable burning 
\citep[][]{2017A&A...604A..48S,2020A&A...636A..77S}. 
This type of outburst shows 1--3\,mag brightening(s) 
in the light curves on a timescale of months to years or even decades 
\citep[e.g.,][]{1991BCrAO..83..104B,
                2005A&A...440..239B,
                2008MNRAS.385..445L,
                2019CoSka..49...19S}
and signs of enhanced mass outflow in the spectrum 
\citep[e.g.,][]{1995ApJ...442..366F,
                2006A&A...457.1003S,
                2011PASP..123.1062M}. 
Stages with Z~And-type outbursts are usually called the active 
phases of SySts. 

During outbursts, an expanding neutral disk-like structure 
emerges around the burning WD in the orbital plane. Its 
presence, being connected exclusively with active 
phases, is transient \citep[][]{2005A&A...440..995S}. 
The disk can be formed from the compression of the enhanced 
wind from the burning WD toward the equatorial plane due 
to the WD rotation \citep[][]{2012A&A...548A..21C}. 
Consequently, the disk blocks ionizing radiation from the central 
burning WD in the orbital plane, which allows the giant’s 
wind to be neutral there. This transient emergence 
of a neutral wind region in the orbital plane during 
outbursts provides us a unique opportunity to measure the H$^0$ 
column densities from the observer to the WD 
($N_{\rm H}^{\rm obs}$) at any orbital phase of eclipsing 
symbiotic binaries \citep[see][]{2023AJ....165..258S}. 

Accordingly, in this work we model the $N_{\rm H}^{\rm obs}$ 
values for all suitable eclipsing SySts during active phases 
and around their orbit measured by 
\cite{2023AJ....165..258S}. 
Following the method of \cite{1993A&A...274.1002K} and 
\cite{2016A&A...588A..83S}, we confirm the substantial focusing of 
the giant's wind toward the orbital plane (Sect.~\ref{s:method}), 
and thus the efficient wind-mass transfer that can operate in 
S-type SySts (Sect.~\ref{s:results}). 
We discuss our results in Sect.~\ref{s:discuss} and present our 
conclusions and suggestions for future work in Sect.~\ref{s:concl}. 

This work is a continuation of our previous articles on the focusing of wind 
from the giant toward the orbital plane in S-type SySts 
(see \citealt{2015A&A...573A...8S}, hereafter Paper~I; 
\citealt{2016A&A...588A..83S}, Paper~II; and 
\citealt{2021A&A...646A.116S}, Paper~III). 
Our efforts are aimed at explaining the large mass accretion rate 
through the stellar wind in these systems, which is needed to explain 
the very high luminosities of their burning WDs 
(Sect.~\ref{ss:dis1}). 

\section{The method}
\label{s:method}
\subsection{Components of $N_{\rm H}^{\rm obs}$}
\label{ss:comps}
Measured values of $N_{\rm H}^{\rm obs}$ consist of three 
components. 
The interstellar component, $< 2\times 10^{21}$\,cm$^{-2}$, 
is estimated from color excesses, $E_{\rm B-V}$, of the used 
targets \citep[see Table~1 of][]{2023AJ....165..258S} 
according to the relationship 
$N_{\rm H}/E_{\rm B-V} \sim 4.93\times 
10^{21}{\rm cm^{-2}mag^{-1}}$ \citep[][]{1994ApJ...427..274D} 
and two circumstellar components. 
The latter are given by the compressed neutral wind from the 
WD during outbursts, $N_{\rm H}^{\rm WD}$, and the neutral 
wind from the red giant (RG), $N_{\rm H}^{\rm RG}$, both 
of which create the neutral near-orbital-plane region as described by 
\cite{2023AJ....165..258S} and shown here in Fig.~\ref{fig:nhfi}. 
According to the simplified model of \cite{2012A&A...548A..21C}, 
the $N_{\rm H}^{\rm WD}$ component is assumed to be constant 
along the orbit, its value is on the order of $10^{22}$\,cm$^{-2}$ 
(from the observer to the WD's pseudo-photosphere of $\sim$20\ro; see 
their Fig.~4), and it also includes the preignition accretion 
disk material. The $N_{\rm H}^{\rm RG}$ component varies within 
two orders of magnitude, dominates all components, and is asymmetric 
with respect to the binary axis (see Fig.~\ref{fig:nhfi}). 
In modeling $N^{\rm obs}_{\rm H}$, we neglected the interstellar 
component because of its relatively very small quantity, and 
$N_{\rm H}^{\rm WD}$ is a variable in our modeling 
(see Eq.~(\ref{eq:nHnum})). 
%
%
%
\begin{table*}[p!t]
\normalsize
\caption[]{
Resulting parameters of modeling the $N_{\rm H}^{\rm obs}$ 
values via function (\ref{eq:nHnum}). 
            }
\begin{center}
\begin{tabular}{cccccccc|cccc}
\hline
\hline
\noalign{\smallskip}
$i$                        & 
$E/I^{(a)}$                & 
$n_1$                      & 
$n_{\rm K}$                & 
$K$                        & 
$N_{\rm H}^{\rm WD}$       & 
$\dot{M}_{\rm sp}^{(b)}$   & 
$\chi^2_{\rm red}{}^{(c)}$ & 
Object                     & 
$d$                        & 
$\dot{M}$                  & 
Ref.                      \\ 
\noalign{\smallskip}
\hline
\noalign{\smallskip}
                     & 
E                    & 
$9.20_{-3.68}^{+3.22}$   & 
$10.9_{-6.5}^{+27.3}$    & 
$6_{-1}^{+1}$            & 
                     &
$2.84_{-1.14}^{+0.99}\times 10^{-6}$ & 
                     &
BF~Cyg               &
3.4                  &
$<3.3\times 10^{-7}{}^{(d)}$ &
(1)                  \\[-1ex]   
\raisebox{1.5ex}{$70^{\circ}$}  & 
I                    & 
$2.20_{-1.10}^{+1.54}$                 & 
$15.7_{-7.9}^{+18.8}$                 & 
$6_{-0}^{+1}$                    & 
\raisebox{1.5ex}{$0.5_{-0.5}^{+0.9}$} &
$6.80_{-3.40}^{+4.76}\times 10^{-7}$  &
\raisebox{1.5ex}{0.47} &
CI~Cyg                 &
1.6                    &
$\sim4.4\times 10^{-7}$& 
(2)                    \\[1ex] 
                     & 
E                    & 
$7.26_{-2.54}^{+4.36}$  & 
$1.33_{-0.80}^{+2.66}$  & 
$4_{-1}^{+1}$           & 
                     &
$2.24_{-0.78}^{+1.34}\times 10^{-6}$ &
                     &
YY~Her               &
6.3                  &
$<2.9\times 10^{-7}$ &
(3)                  \\[-1ex] 
\raisebox{1.5ex}{$80^{\circ}$} & 
I                    & 
$2.05_{-1.13}^{+2.26}$    & 
$3.44_{-1.72}^{+5.16}$    & 
$5_{-0}^{+1}$             & 
\raisebox{1.5ex}{$1.0_{-0.8}^{+1.5}$} &
$6.33_{-3.65}^{+7.29}\times 10^{-7}$ &
\raisebox{1.5ex}{0.35}&
AR~Pav                &
4.9                   &
--                    & 
(4)                  \\[1ex] 
                     & 
E                    & 
$3.84_{-2.69}^{+4.61}$        & 
$0.682_{-0.341}^{+0.750}$     & 
$3_{-0}^{+1}$                 & 
                     &
$1.19_{-0.83}^{+1.43}\times 10^{-6}$ &
                     &
AX~Per               &
1.7                  &
$\sim8.8\times 10^{-8}$ &  
(5)                  \\[-1ex]  
\raisebox{1.5ex}{$90^{\circ}$} & 
I                    & 
$1.79_{-1.43}^{+2.33}$    & 
$1.41_{-0.71}^{+1.55}$    & 
$4_{-0}^{+1}$             & 
\raisebox{1.5ex}{$1.5_{-1.2}^{+1.5}$} &
$5.53_{-4.24}^{+7.19}\times 10^{-7}$ &
\raisebox{1.5ex}{0.43} &
PU~Vul                 &
4.7                    &  
$<5.8\times 10^{-7}{}^{(d)}$ & 
(6)                   \\
\noalign{\smallskip}
\hline
\end{tabular}
\end{center}
{\bf Notes:} 
The first part of the table contains the fitting parameters of 
the wind, $n_1 (10^{23}$\,cm$^{-2})$, 
$n_{\rm K} (10^{25}$\,cm$^{-2})$, $K$, 
$N_{\rm H}^{\rm WD}(10^{22}$\,cm$^{-2})$, 
the corresponding spherical equivalent of the mass-loss rate, 
$\dot{M}_{\rm sp}$ ($M_{\odot}\,{\rm yr}^{-1}$; see 
Sect.~\ref{ss:res1}), and a minimum of the $\chi_{\rm red}^2$ 
function for three different orbital inclinations, $i$ (see 
Sect.~\ref{ss:modeling}). 
The second part, to the right of the vertical bar, compares the total 
mass-loss rates, $\dot{M}$ ($M_{\odot}\,{\rm yr}^{-1}$), 
derived from radio emission by \cite{1993ApJ...410..260S} 
recalculated for distances $d$ (in kpc) from the literature, 
listed in the last column, Ref. 
%
$^{(a)}$\,From egress (E) or ingress (I) data 
(see Sect.~\ref{ss:modeling}). 
$^{(b)}$\,For $v_\infty = 30$\,km\,s$^{-1}$ 
(see Sect.~5.1. of Paper~II). 
$^{(c)}$\,Corresponds to 34 degrees of freedom. 
$^{(d)}$\,Upper limit due to a contribution from the active 
hot component. References: 
(1) \cite{1991A&A...248..458M}, 
(2) \cite{1993ApJ...410..260S},
(3) \cite{2005A&A...440..995S},
(4) \cite{2001A&A...366..972S},
(5) \cite{2001A&A...367..199S},
and (6) \cite{2012ApJ...750....5K}.
\label{t:results}
\end{table*}

\subsection{Modeling $N_{\rm H}^{\rm obs}$ along the orbit}
\label{ss:modeling}
\subsubsection{The principle of the method}
\label{sss1:modeling}
Using the equation of continuity, the theoretical value of 
the total hydrogen column density of the spherically symmetric wind 
from the giant, $\tilde{N}_{\rm H}$, along the line of sight from 
the observer ($-\infty$) to infinity that contains the WD, $l,$ can 
be expressed as 
\begin{equation}
  \tilde{N}_{\rm H} = \displaystyle\frac{\dot M}{4\pi \mu m_{\rm H}}
               \displaystyle\int\limits_{-\infty}^{\infty}
               \displaystyle\frac{{\rm d}l}{r^2v(r)}, 
\label{eq:nH}
\end{equation}
where $\dot M$ is the mass-loss rate from the RG, $\mu$ the mean 
molecular weight, $m_{\rm H}$ the mass of the hydrogen atom, 
$r$ the distance from the RG center, and $v(r)$ the wind velocity 
profile (WVP; which is the velocity of the wind particles at 
a radial distance from the giant's center). 

The form of integral (\ref{eq:nH}) allows the application of 
the inversion method via the diagonalization of the integral operator 
for column density \citep[see][]{1993A&A...274.1002K}. This approach 
yields three essential relationships (see Paper~II): 
\begin{enumerate}
\item
The parameterized total hydrogen column density of the giant's 
wind from the inversion method can be expressed 
as \citep[][]{1999A&A...349..169D} 
\begin{equation}
\label{eq:nHb}
   \tilde{N}_{\rm H}(b) = \frac{n_1}{b} + \frac{n_K}{b^K},
\end{equation}
where $n_1$, $n_K$, and $K$ are fitting parameters, and $b$ is 
the separation between the binary components, $p$, projected onto 
the plane perpendicular to the line of sight, the so-called 
impact parameter: $b^2=p^2(\cos^2 i + \sin^2 \varphi \sin^2 i)$, 
where $i$ and $\varphi$ are the orbital inclination and 
the orbital phase \citep[see][]{1991A&A...249..173V}. 
\item
The WVP, 
\begin{equation}
\label{eq:wvp}
   v(r) = \frac{v_\infty}{1 + \xi r^{1-K}},
\end{equation}
where $v_\infty$ is the terminal velocity of the wind, and 
$\xi = {n_K\lambda_1}/{n_1\lambda_K}$ is the parameter 
of the wind model, where $\lambda_1 = \pi/2$ and 
$\lambda_i = \lambda_1/(i-1)\lambda_{i-1}$ ($i\geq 2$) 
are the eigenvalues of the Abel operator 
\citep[][]{1993A&A...274.1002K}. 
The parameter $\xi$ determines the distance from the RG surface 
at which the wind starts to accelerate significantly, while 
the parameter $K$ defines the steepness of this acceleration 
(see Paper~II). 
\item
The mass-loss rate \citep[see][]{2023A&A...676C...3S}, 
\begin{equation}
 \dot{M} = 2\pi\mu m_{\rm H}R_{\rm G}
           \frac{n_1}{\lambda_1}v_\infty ~~ ({\rm g\,s^{-1}}) . 
\label{eq:dotM}
\end{equation}
\end{enumerate}
In our approach, the asymmetric course of $N^{\rm obs}_{\rm H}$ 
values relative to the binary axis (see Fig.~\ref{fig:nhfi}) 
constrains two different WVPs -- the ingress one within 
$\varphi\sim 0.75 - 1$, which represents the hydrogen density distribution 
in front of the RG orbital motion, and the egress profile within 
$\varphi\sim 0 - 0.25$ resulting from the density distribution 
behind the RG motion. 
For the orbital phases between these ranges, we assumed that 
the velocity profile at the plane of observations changes 
gradually in a smooth way from the egress to the ingress profile. 
We used an interconnection method, which fulfills the conditions 
of smoothness for both the velocity profiles and the column 
density as functions of the orbital phase 
\citep[see][]{2017A&A...602A..71S}. 
Accordingly, we calculated the column density profile of 
the neutral near-orbital-plane region from the observer 
to the WD as \citep[see][]{2017A&A...602A..71S}
\begin{equation}
  N_{\rm H}(b) = \frac{\dot{M}}{4\pi \mu m_{\rm H}}
                 \int\limits_{-\infty}^{\pm\sqrt{p^2-b^2}}
                 \frac{dl}{(l^2+b^2)v(\sqrt{l^2+b^2})} + 
                 N_{\rm H}^{\rm WD}, 
\label{eq:nHnum}
\end{equation}
where the first term on the right represents the column density 
of the RG wind, $N_{\rm H}^{\rm RG}$, while 
the second one, $N_{\rm H}^{\rm WD}$, is the contribution of 
the compressed neutral wind from the WD with the preignition 
accretion disk between the WD's warm pseudo-photosphere and 
the observer (see Sect.~\ref{ss:comps}). 
The term $v(\sqrt{l^2+b^2})$ represents the egress, ingress, 
or transitional wind velocity, depending on the location 
on the line of sight and the orbital phase. 
The upper limit of integration corresponds to the position of 
the WD, where the usage of the plus or minus sign depends on 
the position of the binary, that is, on the angle between the 
line of sight and the binary axis 
(see \citealt{2017A&A...602A..71S} for details). 
For the binary parameters, $p$ and $R_{\rm G}$, we adopted 
typical values of 400\ro\ and 100\ro, respectively 
\citep[][]{2000A&AS..146..407B}. 
Finally, the variables determining the model 
$N_{\rm H}(b)$ (Eq.~(\ref{eq:nHnum})) are the $n_1$, $n_K$, and $K$ 
of the WVP (Eq. (\ref{eq:wvp})), the value of $N_{\rm H}^{\rm WD}$ 
(Sect.~\ref{ss:comps}), and the corresponding mass-loss rate, 
$\dot{M}$ (Eq.~(\ref{eq:dotM})). 

\subsubsection{Description of the fitting procedure}
\label{sss2:modeling}
We determined the model parameters by fitting the 
$N^{\rm obs}_{\rm H}$ values with function (\ref{eq:nHnum}) 
using our own software\footnote{The Fortran codes and 
an application example are available at 
https://zenodo.org/record/8120552}. 
To obtain the resulting parameters, we first estimated possible 
initial ranges of the WVP parameters $n_1$, $n_K$, and 
$K$ \footnote{In the first step, we chose relatively large ranges, 
$10^{20} - 10^{30}$\,cm$^{-2}$, $10^{20} - 10^{40}$\,cm$^{-2}$, 
and $1 - 40$ for the parameters $n_1$, $n_K$, and $K$, respectively, 
to avoid missing the best solution.} while keeping 
$N_{\rm H}^{\rm WD}$ \footnote{We selected $N_{\rm H}^{\rm WD}$ 
values from the interval $0 - 10^{23}$\,cm$^{-2}$.} fixed. We iterated 
the following procedure: 
\begin{enumerate}
\item
We randomly generated the values of  all ingress and egress WVP 
fitting parameters. 
\item
We integrated column densities (see Eq.~(\ref{eq:nHnum}) for the 
corresponding WVPs using the fourth-order Runge-Kutta method. 
\item
We evaluated the goodness of fits via the least-square method. 
\end{enumerate}
Overall, we calculated 50\,000 models for random WVP parameter 
values in the first run. Using the top 50 models, we 
determined their new narrower ranges. 
We repeated this procedure until the desired accuracy of 
the resulting WVP parameters (approximately a few percent) was 
achieved. 

\begin{figure}
\begin{center}
\resizebox{\hsize}{!}
          {\includegraphics[angle=-90]{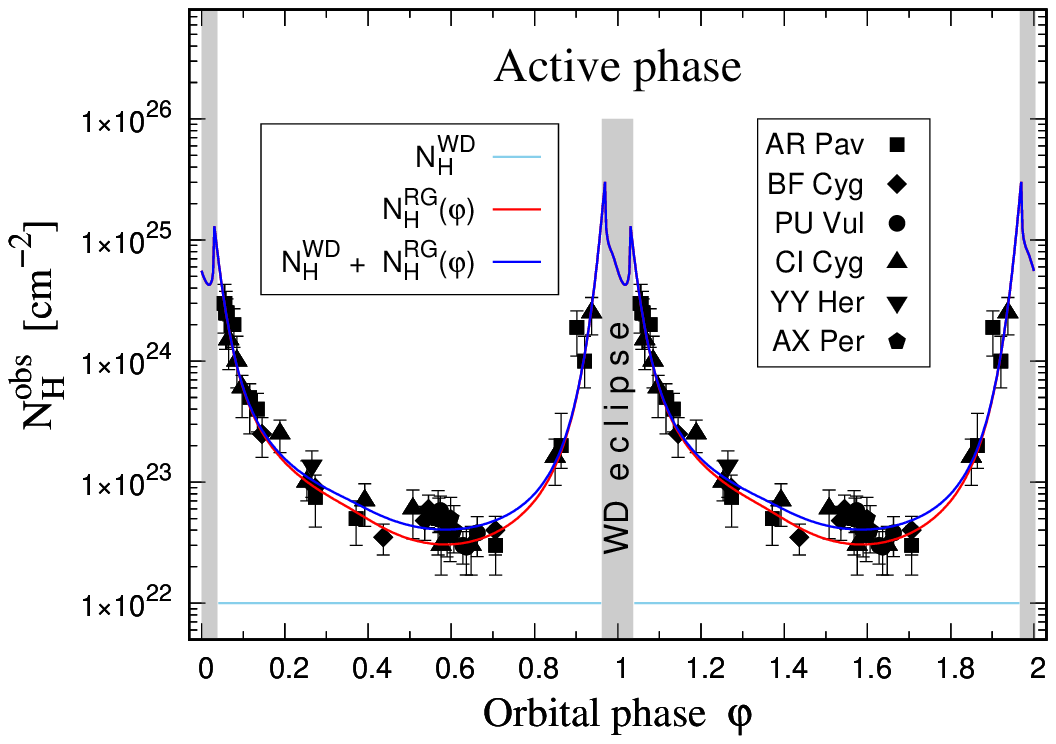}}
\end{center}
\caption[]{
Column densities of atomic hydrogen between the observer 
and the WD ($N_{\rm H}^{\rm obs}$) measured for eclipsing 
SySts during active phases around the whole orbit 
\citep[data from][]{2023AJ....165..258S}. 
The blue line is the best-fitting model for $i=80^{\circ}$ 
(Table~\ref{t:results}). 
It consists of a constant value along the orbit, 
$N_{\rm H}^{\rm WD} = 1\times 10^{22}$\,cm$^{-2}$, 
supplied by the compressed neutral wind from the WD during 
outbursts, and the phase-dependent value, 
$N_{\rm H}^{\rm RG}(\varphi)$, of the neutral wind from 
the RG (see Sect.~\ref{ss:comps}). 
The minimum in the model during the WD eclipse is caused by 
the fact that the line of sight ends at the surface 
of the RG, and not on the warm WD's pseudo-photosphere. 
Asymmetrical distribution and high values of 
$N_{\rm H}^{\rm obs}$ reflect the asymmetric wind from 
the RG and its focusing on the orbital plane 
(see the main text). 
          }
\label{fig:nhfi}
\end{figure}

For the given $N_{\rm H}^{\rm WD}$ and the best corresponding 
ingress and egress 
models, we compared the column densities 
given by Eq.~(\ref{eq:nHnum}) with the measured 
$N_{\rm H}^{\rm obs}$ values at all orbital phases. In this 
way, varying the $N_{\rm H}^{\rm WD}$ parameter, we selected 
the best solution. 
The resulting parameters, $n_1$, $n_{\rm K}$, and $K$ (which 
define the WVP), the column density, $N_{\rm H}^{\rm WD}$, and 
the mass-loss rate, $\dot{M}$, are listed in Table~\ref{t:results}, 
and Fig.~\ref{fig:nhfi} shows the model for $i=80^{\circ}$. 

Finally, we note 
that due to the gap in the data set around $\varphi = 0.8$, 
the least-square method did not always provide adequate values 
of the ingress $n_1$ parameter. In these cases, we estimated 
its value so as to get a reasonable model here: For $i=70^{\circ}$, 
we obtained the appropriate value of the ingress $n_{1}$ parameter 
by fixing its value to $2.2\times 10^{23}$\cmd\ and performing 
steps 1 to 3 of the fitting procedure. For $i=80^{\circ}$, we 
fixed $n_1 = 2.05\times 10^{23}$\cmd\ and $K=5$. For $i=90^{\circ}$; 
no manual intervention was needed. 

\subsubsection{Determination of uncertainties}
\label{sss3:modeling}
Computing a grid of models by individually varying the values of 
fitting parameters around the best solutions, we estimated their 
uncertainties given by the errors and/or spread in 
$N^{\rm obs}_{\rm H}$ values. 
Typically, these errors are large ($\gtrsim 50\%$ and higher), 
with the upper uncertainty larger than the lower. This is also true 
for the resulting values of the $\dot{M}_{\rm sp}$, with average 
uncertainties at the level of $\sim 50\%$ for the lower limit 
and $\sim 90\%$ for the upper limit. The average value of 
$\dot{M}_{\rm sp}$ over all considered inclinations and 
for both the egress and ingress cases is 
$1.36\times 10^{-6}\,M_{\odot}\,{\rm yr}^{-1}$ with a standard 
deviation of $9.63\times 10^{-7}\,M_{\odot}\,{\rm yr}^{-1}$ 
(see Table~\ref{t:results}). We also note that the uncertainty 
in the $\dot{M}_{\rm sp}$ linearly reflects the uncertainty in 
$v_{\infty}$ (see Eq.~(\ref{eq:dotM})). 

The uncertainties of the $N^{\rm WD}_{\rm H}$ values 
are estimated from the $N_{\rm H}(b)$ profile 
around $\varphi = 0.5$, where the RG wind is sufficiently 
thin along the line of sight to allow us to identify the effect 
of the $N^{\rm WD}_{\rm H}$ contribution. 
The large uncertainty of the ingress $n_1$ parameter is directly 
connected to the uncertainty of the $N^{\rm WD}_{\rm H}$ values 
shown in Table~\ref{t:results}. 

Although all the used objects are eclipsing binaries, their 
accurate $i$ are not known. Therefore, we performed 
modeling for $i=90^{\circ}$, $80^{\circ}$, and 
$70^{\circ}$ to demonstrate the effect of an uncertainty in 
$i$ on the model parameters. For example, lower values of $i$ 
require higher values of $\dot{M}$ to achieve good agreement 
with the measured column densities. 
Similarly, the column density given by Eq.~(\ref{eq:nHb}), as well 
as the difference between the egress 
and ingress data, decreases with $i$ (Table~\ref{t:results}). 
Therefore, the model requires a higher column density asymmetry 
in the ingress and egress orbital phases for lower $i$. 
Values of the parameter $n_{\rm K}$, which reflects the density 
conditions close to the RG surface, are a strong function 
of $i$, given by the $K$-th power of the impact parameter $b$ 
in the second term of Eq.~(\ref{eq:nHb}). On the contrary, the 
values of the parameter $n_1$, which represents the density 
farther away from the RG, do not show significant variability 
with $i$. This also reflects an insignificant dependence of 
$\dot{M}$ on $i$ (see Eq.~(\ref{eq:dotM}) and 
Table~\ref{t:results}). 
%
%
%
%

\subsection{Applicability of the method}
\label{ss:appl}
Observability of the Rayleigh scattering effect in a binary, and 
thus the application of our method, does not depend on the separation 
between the accreting WD and its donor. 
The method just requires the line of sight to pass through 
a measurable amount of the neutral wind from the giant that we 
can model. The ionization structure 
restricts the application of the method to objects with a high 
$i$ \citep[see Fig.~2f of][]{2023AJ....165..258S}. 
During quiescent phases, the number of H$^0$ atoms 
is measurable only around the inferior conjunction of the RG, 
from the observer to the H$^0$/H$^+$ boundary, while during 
outbursts we can measure its quantity from the observer to 
the WD and around the whole binary. 
The widest symbiotic binary for which we have the required 
observations is PU~Vul 
\citep[orbital period of 13.4 years; see][]{2012BaltA..21..150S}. 

\section{Results}
\label{s:results}
\subsection{Enhanced wind-mass loss in the orbital plane}
\label{ss:res1}
Our column density models of the neutral near-orbital-plane 
region (Eq.~(\ref{eq:nHnum})) correspond to the mass-loss rate 
from the giant, $\dot{M}\approx 10^{-6}$\myr\ 
(see Table~\ref{t:results}). 
However, a mass-loss rate of $\approx$10$^{-7}$\myr\ 
has been estimated from the nebular emission of the ionized 
wind from the giant during quiescent phases of S-type SySts 
\citep[see][]{1991A&A...248..458M,
      1993ApJ...410..260S,
      2002AdSpR..30.2045M,
      2005A&A...440..995S}. 
Because the latter approach does not depend on the line 
of sight (above all, it is independent of $i$), 
the corresponding $\dot{M}$ can be considered the total 
mass-loss rate. 

The difference between our value of $\approx$10$^{-6}$\myr\ 
and the total value of $\approx$10$^{-7}$\myr\ may be 
caused by the non-spherically symmetric distribution of 
the particle density in the wind. 
In our case, using $N_{\rm H}^{\rm obs}$ from 
the near-orbital-plane region only, we can determine only 
the spherical equivalent of the mass-loss rate, 
$\dot{M}_{\rm sp}$, which assumes a spherically 
symmetric distribution of the particle density in the wind. 
If this were the case, then the equality 
$\dot{M}_{\rm sp} = \dot{M}$ would apply. 
Otherwise, the  $N_{\rm H}^{\rm obs}$ from denser parts of 
the wind yields $\dot{M}_{\rm sp} > \dot{M}$, and vice versa. 
Thus, comparing the total value of the mass-loss rate with 
the $\dot{M}_{\rm sp}$ can give us important information 
about the density structuring of the wind. 
In our case of solely eclipsing systems, the line of sight 
pointing to their WDs always passes through the near-orbital-plane 
region. Therefore, our finding that $\dot{M}_{\rm sp}>\dot{M}$ 
(Table~\ref{t:results}) implies enhanced wind-mass loss in 
the orbital plane. This result is consistent with that found 
for the quiescent SySts EG~And and SY~Mus (see Paper~II). 

\subsection{High wind-mass-transfer efficiency}
\label{ss:res2}
The increase in the wind-mass-loss rate in the orbital 
plane with the factor $\sim\dot{M}_{\rm sp}/\dot{M}$ implies 
a proportional increase in the accretion rate with respect 
to the spherically symmetric case. 
Hence, this result points to a higher efficiency of the 
wind-mass transfer between the binary components than what 
is expected for the canonical Bondi-Hoyle type of wind 
accretion \citep[][]{1944MNRAS.104..273B}. 
Our finding, which is derived from observations, is consistent 
with theoretical calculations of gravitational enhancement of 
the wind-mass loss in the orbital plane due to the presence of 
a binary companion \citep[][]{
2001A&A...367..513F,
2009ApJ...700.1148D,
2017MNRAS.468.3408D,
2020MNRAS.493.2606B,
2020A&A...637A..91E}, 
which can be enhanced by the compression of the wind toward the 
orbital plane due to the rotation of normal giants in S-type 
SySts (see Paper\,I and references therein). 
Also, the indication of higher H$^0$ column densities between 
the binary components compared to opposite orbital phases, 
recently revealed for the eclipsing SySt EG~And by 
\cite{2023A&A...676A..98S}, is likely the result of enhanced 
mass transfer between binary components. 
Here, our indication of a high wind-mass-transfer efficiency 
is consistent with our previous findings of the wind focusing 
toward the orbital plane (see Papers I, II, and III). 

\subsection{Confirmation of the wind asymmetry}
\label{ss:res3}
\begin{figure}
\begin{center}
\resizebox{\hsize}{!}
          {\includegraphics[angle=-90]{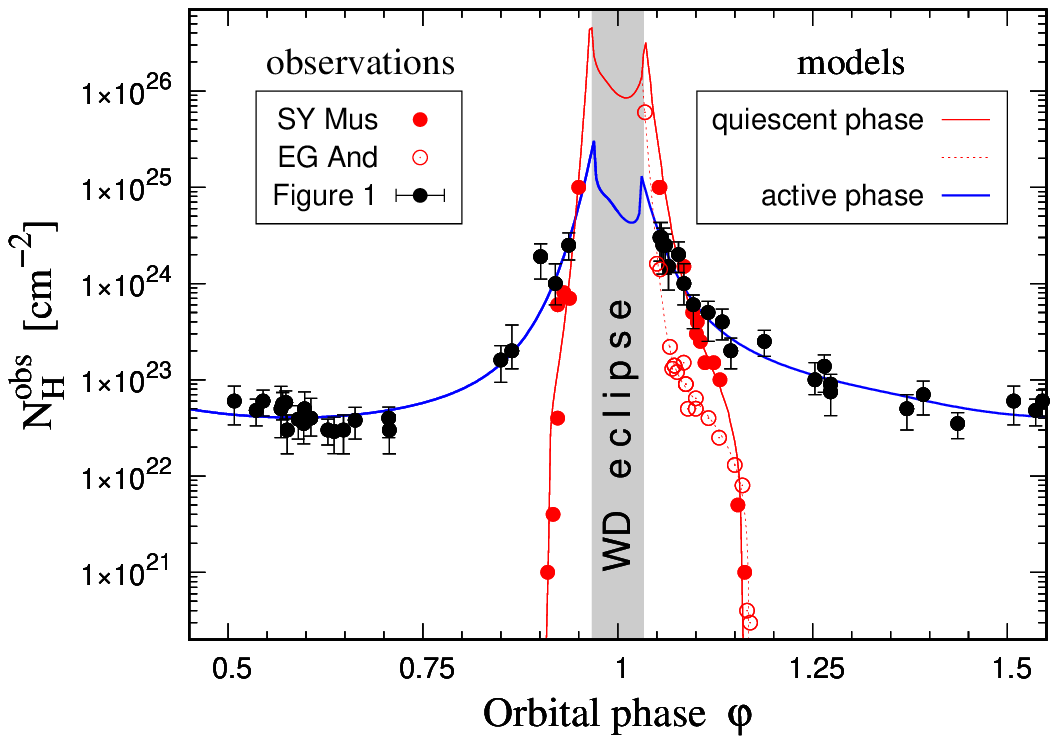}}
\end{center}
\caption[]{
Comparison of the $N_{\rm H}^{\rm obs}$ column densities for 
eclipsing systems (i.e., measured in the orbital plane) during 
quiescent phases (red symbols; data and models from Paper~II) 
and active phases (black circles; data and the model from 
Fig.~\ref{fig:nhfi}). 
          }
\label{fig:nhqa}
\end{figure}

Asymmetric distribution of wind particle density in the orbital 
plane with respect to the binary axis (see Fig.~\ref{fig:nhfi} 
and Sect.~3.3 of \citealt{2023AJ....165..258S}) 
constrains the asymmetric WVP because the velocity of the wind 
determines the function for column densities (Eq.~(\ref{eq:nHnum})). 
The ingress of the WVP being steeper than its egress (Eq.~(\ref{eq:wvp}) 
and Table~\ref{t:results}) implies a higher density gradient in 
front of the RG orbital motion than behind it. 
During quiescent phases, this manifests itself in the asymmetry 
of the ionization boundary in the orbital plane with respect to 
the binary axis \citep[see Fig.~2c of][]{2023AJ....165..258S} 
and in the asymmetry of the far-UV light curves around the zero 
orbital phase \citep[see][]{2017A&A...602A..71S}. 
The former is given by the asymmetric distribution of 
$N_{\rm H}^{\rm obs}$ values around the inferior conjunction of 
the RG (see Fig.~\ref{fig:nhqa}), while the latter results from 
the different attenuation of the WD's radiation along 
the directions in front of and behind the RG orbital motion.

As the wind asymmetry alongside the orbital motion of the RG is 
indicated during both the quiescent and active phases, and for 
different objects, this apparently represents a common feature 
of the wind material from normal giants in symbiotic 
binaries. 
%
%
%

\section{Discussion}
\label{s:discuss}
\subsection{Wind-focusing as a new mass-transfer mode 
            in S-type SySts}
\label{ss:dis1}
The first indication of wind-focusing was suggested for 
the nearby SySt SS~Leporis (distance of 280 parsecs, orbital period 
of 260 days) by interferometric measurements that resolved both 
binary components. The measurements revealed the RG to be almost 
twice smaller than its Roche lobe, which allows one to conclude 
that an enhanced wind-mass loss in the orbital plane fuels 
the abnormal luminosity of the hot component 
\citep[see][]{2011A&A...536A..55B}. 
%
For the quiet eclipsing symbiotic systems EG~And and SY~Mus, 
we determined the wind-focusing toward the orbital plane by 
modeling the $N_{\rm H}^{\rm obs}$ column densities around 
the inferior conjunction of the RG (see Paper~II and  our
Fig.~\ref{fig:nhqa}), which revealed a mass-loss rate in the orbital plane that is a factor of 
$\sim$10 higher than 
its total value. 
%
Recently, we indirectly confirmed the compression of the giant's 
wind in the orbital plane for EG~And by discovering significant 
dilution around the poles of the giant that corresponds to 
$\dot{M}_{\rm sp}\le 10^{-8}$\myr\ (see Paper~III). 
%
In the present study we confirm the wind-focusing toward 
the orbital plane for eclipsing SySts during active phases. 
Wind-focusing for quiescent systems, both 
the accretion-powered SySt EG And and the nuclear-powered 
SY~Mus, as well as for systems transiting the active phase, 
suggests that this property may be a common feature of the wind 
from normal giants in SySts. 

The abovementioned findings indicate the presence of an 
efficient wind-mass-transfer mode operating in S-type SySts, 
especially those with nuclear shell burning, in which the high 
luminosity of burning WDs requires a high accretion rate. 
An effective wind-mass-transfer mode is a key toward 
resolving a long-standing fundamental problem in SySts research: 
the discrepancy between the large energetic output from 
burning WDs and their deficient fueling by the giant 
in the Bondi-Hoyle type of wind accretion
\citep[first pointed by][]{1983AJ.....88..666K} because of its 
low efficiency\footnote{Defined as the mass accretion rate divided 
by the mass-loss rate.} of only 0.6 to 10\% \citep[see][]{
1944MNRAS.104..273B,
1996MNRAS.280.1264T,
2004A&A...419..335N,
2013ApJ...764..169P}. 
In particular, the high luminosities of nuclear-powered 
systems require a high accretion rate of 
$\sim(10^{-8}-10^{-7}$)\myr\ \citep[e.g.,][]{1978ApJ...222..604P,
2007ApJ...660.1444S}, which cannot 
be fueled via the Bondi-Hoyle accretion because the mass-loss 
rate from RGs in S-type systems is on the same order 
($\approx10^{-7}$\myr; Sect.~\ref{ss:res1}). 

Another important implication of wind-focusing is the fueling 
of the high-mass WDs in recurrent symbiotic novae via the wind, 
which also requires accretion rates of a few times 
10$^{-8}$\myr\ onto a 1.25--1.40\mo\ WD \citep[e.g.,][]{
2005ApJ...623..398Y,
2009ApJ...697..721S,
2016ApJ...825...95D,
2021MNRAS.501..201H}. 

\subsection{Wind-focusing and ellipsoidal light variations}
\label{ss:dis2}
Due to the large separation of the binary components in SySts, 
the mass from the RG is expected to be transferred onto its 
compact companion via the stellar wind 
\citep[e.g.,][]{1999A&AS..137..473M}. 
However, the light curves of many SySts show so-called 
ellipsoidal variation \citep[i.e., two minima and two maxima per 
orbital cycle; see, e.g.,][]{2013AcA....63..405G}, which is 
generally understood to be a result of the Roche-lobe filling giant. 
Therefore, this effect has been considered an indication of 
the mass-transfer via Roche-lobe overflow (RLOF) for 
symbiotic binaries 
\citep[see][]{1997MNRAS.291...54W,
2002A&A...392..197M,
2003ASPC..303..151M}. 
However, there are a number of arguments against this natural 
interpretation. We summarize them as follows. 
\begin{enumerate}
\item
Modeling the corresponding optical and near-IR light curves 
shows that the required Roche-lobe filling factor, 
$R_{\rm RG}/R_{\rm L}$, is inconsistent with the cool 
component radii derived from measured v$\sin(i)$ and spectral 
type for most systems 
\citep[][]{
2007BaltA..16...49R,
2007BaltA..16....1M,
2012BaltA..21....5M,
2014ASPC..490..367O}. 
\item
The use of other methods also shows that the actual radii of 
RGs in SySts are too small to fill their Roche lobes. 
For example, (i) \cite{1992A&A...260..156V} determined the 
radius of an M2.4\,III giant in EG~And ($P_{\rm orb} = 482.2$\,d) 
to be only 75\ro, from the 1991 eclipse observed in 
the UV. 
(ii) Using a relation between the visual surface brightness 
and the Cousins $V-I$ color index, \cite{1998NewA....3..137D} 
found that radii of normal M giants increase from a median value 
of 50\ro\ at spectral type M0\,III to 170\ro\ at M7/M8\,III, 
for available Hipparcos parallaxes. 
(iii) Based on a strong correlation between the spectral type 
of the RG and the orbital period for 30 SySts, 
\cite{1999A&AS..137..473M} found that the RG radii 
$R_{\rm RG}\le l_{1}/2$ with only one exception (T~CrB), 
where $l_{1}$ is the distance from the center of the RG 
to the inner Lagrangian point. 
(iv) From the spectral energy distribution of RGs in S-type 
SySts and given distances, \cite{2005A&A...440..995S} found 
that their radii are far shorter than their Roche-lobe radii 
(see Tables 1 and 2 therein). 
(v) The symbiotic binary V1261~Ori shows a pronounced ellipsoidal 
variation in its ASAS $V$-band 
light curve, but the interferometric measurements suggest its 
Roche-lobe filling factor to be only $\sim$0.3 
\citep[see][]{2014A&A...564A...1B}. 
\item
The largest discrepancy between the true radii of RGs and their 
Roche-lobe radii in symbiotic binaries is most apparent for some 
yellow SySts, whose optical light curves show marked ellipsoidal 
variation \citep[see Fig.~6.2 of][]{2019arXiv190901389M}, although 
their G-K giants with radii typically of 30--50\ro\ are deep 
inside their Roche lobes. 
\item
Finally, our analysis also favors wind-mass transfer over RLOF
because the density profile in Fig.~\ref{fig:nhfi} corresponds 
to the WVP (Eq.~(\ref{eq:wvp})). 
\end{enumerate}
The above arguments indicate that the nature of the ellipsoidal 
variation in the light curves of some SySts differs from that of 
the tidally distorted cool giants that fill their Roche lobe. 
%
Here, we propose that 
the substantial focusing of the wind toward the orbital plane 
could mimic the tidally distorted cool giants and thus produce 
the ellipsoidal variation in the light curves, even for systems 
containing RGs well within their Roche lobes. 
However, this idea needs further quantitative confirmation, which 
is beyond the scope of this paper. 

\subsection{Comparison with D-type SySts}
\label{ss:dis3}
The efficient mass-transfer mode in D-type systems is conditioned 
by a slowly accelerating wind produced by a Mira-type variable 
with an acceleration radius that lies close to the Roche-lobe radius. 
In this case, the slow wind fills the Mira's Roche lobe instead 
of the star itself and can be effectively transferred through 
the L$_1$ point to the WD -- the so-called wind Roche-lobe 
overflow \citep[WRLOF; see][]{
2007ASPC..372..397M,
2012BaltA..21...88M,
2013A&A...552A..26A}. 
By simulating this for $o$~Ceti 
(with a projected binary separation of $\sim$65\,a.u. and a wind 
velocity of $\approx$5\,km\,s$^{-1}$), \cite{2006ApJ...637L..49M} 
found that the mass transfer efficiencies can be at least an order 
of magnitude higher than the analogous Bondi-Hoyle values. 
Thus, despite the large separation of the binary components, 
the efficiency of the mass transfer in D-type SySts is comparable 
with that in the S-types. 

In principle, our method could also be used to examine the wind 
for D-type systems; however, their orbital parameters are mostly 
unknown, and thus there are no appropriate observations. 
Finally, we note that the WRLOF mode is probably not applicable 
to wind from normal giants in S-type SySts, which is a few times faster  
\citep[][ Paper~II]{2007ASPC..372..397M}, although proper modeling 
should be carried out to obtain a definitive conclusion. 
%

\section{Conclusion and future work}
\label{s:concl}
In this work we modeled the $N_{\rm H}^{\rm obs}$ column 
densities of the wind from normal giants in eclipsing S-type 
SySts during their outbursts (see Sect.~\ref{s:intro}, 
Fig.~\ref{fig:nhfi}, and Eq.~(\ref{eq:nHnum})). 
We find that the spherical equivalent of the mass-loss rate 
derived from the orbital-plane column densities can be a factor 
of $\approx$10 larger than the total value (Sect.~\ref{ss:res1}). 
This implies a focusing of the wind toward the orbital plane 
and, thus, its more efficient transfer onto the WD compared to 
the standard Bondi-Hoyle type of wind accretion 
(Sect.~\ref{ss:res2}). 
Our findings independently confirm the effect of wind-focusing 
previously found for the quiescent SySts SY~Mus and EG~And and 
help us understand the discrepancy between the too high 
luminosity of H-burning WDs and their insufficient fueling 
through canonical Bondi-Hoyle wind accretion 
(Sect.~\ref{ss:dis1}). 

We have several suggestions for future work. The compression 
of the giant's wind toward the orbital plane presents 
an interesting challenge for further theoretical modeling 
of the stellar wind morphology from giants in S-type SySts:

\begin{enumerate}
\item
Modeling of the accretion process that meets the conditions 
derived from observations ($\dot{M}_{\rm sp}/\dot{M}\approx 10$ 
and $<0.1$ in the orbital plane and around the poles of the RG, 
respectively, and a total mass-loss rate of $\dot{M}\approx 10^{-7}$\myr) 
should more accurately determine the mass-transfer mode operating 
in S-type SySts (see Sect.~\ref{ss:dis1}). 
Also, determining the ratio of mass-loss rates, 
$\dot{M}_{\rm sp}/\dot{M}$, as a function of $i$ would 
illustrate the pole--equator asymmetry of the wind material 
from normal giants in SySts. 
\item
Consequently, the implementation of the efficient wind-mass-transfer 
mode in the binary population synthesis codes can improve current 
simulations of the birth rate and number of SySts 
\citep[e.g.,][]{2006MNRAS.372.1389L}. 
\item
Theoretical calculations of the optical depth of the compressed 
material in the orbital plane are needed to test whether the 
corresponding light curves are similar to those produced 
by the tidally distorted giants (see Sect.~\ref{ss:dis2}).
\end{enumerate}

Finally, we note that 
the expected high accretion rates from the wind compressed in 
the orbital plane should allow the WD to grow faster in mass 
to the Chandrasekhar limit despite the hydrogen and helium 
flashes on its surface \citep[see][]{2016ApJ...819..168H} and 
thus explode as a Type Ia supernova. 
Therefore, the wind-focusing toward the orbital plane makes S-type 
symbiotic binaries an effective wind-accretion channel 
for producing Type~Ia supernovae. 
%
\begin{acknowledgements}
We thank the anonymous referee for valuable comments. 
This work was supported by a grant of the Slovak Academy 
of Sciences, VEGA No. 2/0030/21, and by the Slovak Research 
and Development Agency under contract No. APVV-20-0148. 
This research has made use of NASA’s Astrophysics Data System 
Bibliographic Services. 
\end{acknowledgements}


\bibliographystyle{aa}

\bibliography{focus4.bib}

\end{document}